\documentclass[epsfig,10pt,twocolumn]{article}
\usepackage{epsfig,citesort}
      \input{amssym.def}
      \input{amssym}
      \newcommand {\mm}[1] {\ifmmode{#1}\else{\mbox{\(#1\)}}\fi}

\renewcommand{\thetable}{\Roman{table}}

\newcommand{\utwi}[1]{\mbox{\boldmath $ #1$}}

\newcommand{\bx}{{\utwi{x}}}

\begin{document}
      \title{\bf Simplicial edge representation of protein structures and
alpha contact potential with confidence measure}

      \author{\bf Xiang Li, Changyu Hu and Jie Liang\thanks{Corresponding author. Phone:
      (312)355--1789, fax: (312)996--5921, email: {\tt
      jliang@uic.edu}} \\ Department of Bioengineering, SEO, MC-063 \\
      University of Illinois at Chicago\\ 851 S.\ Morgan Street, Room
      218 \\ Chicago, IL 60607--7052, U.S.A.
      \\
      (Accepted by {\it Proteins.})
      }  \date{\today}

      \maketitle

\abstract{ \it Protein representation and potential function are two
 important ingredients for studying proteins folding, equilibrium
 thermodynamics, and sequence design.  We introduce a novel geometric
 representation of protein contact interactions using the edge
 simplices from alpha shape of protein structure.  This representation
 can eliminate implausible neighbors that are not in physical contact,
 and can avoid spurious contact between two residues when a third
 residue is between them.  We develop statistical alpha contact
 potential using an odds-ratio model.  A studentized bootstrap method
 is then introduced for assessing the 95\% confidence intervals for
 each of the 210 propensity parameters.  We found with confidence that
 there is significant long range propensity ($>$ 30 residues apart)
 for hydrophobic interactions.  We test alpha contact potential for
 native structure discrimination using several sets of decoy
 structures, and found it often has comparable performance with
 atom-based potentials requiring many more parameters.  We also show
 that accurate geometric representation is important, and alpha
 contact potential has better performance than potential defined by
 cut-off distance between geometric centers of side chains.
 Hierarchical clustering of alpha contact potentials reveals natural
 grouping of residues.  To explore the relationship between shape
 representation and physicochemical representation, we test the
 minimum alphabet size necessary for native structure discrimination.
 We found that there is no significant difference in performance of
 discrimination when alphabet size varies from 7 to 20, if geometry is
 represented accurately by alpha simplicial edges.  This result
 suggests that the geometry of packing plays an important role, but
 the specific residue types are often interchangeable.  }


\section{Introduction}

Potential function plays important roles for a variety of
computational studies of proteins, including prediction of protein
structures, characterization of ensemble thermodynamic properties of
proteins, and design of novel proteins.  For example, an essential
requirement for the prediction of three-dimensional structure of
protein from primary sequence is a potential function which can select
the native conformation from an ensemble of alternative conformations.
Potential function is also often used to guide the sampling of protein
conformations \cite{Skolnick99_ACP}.  A variety of potential functions
have been developed for these important tasks, including physical
model based potentials
\cite{Jorgensen88_JACS,Lazaridis99_JMB,GratchellVajda00_Proteins},
empirical statistical potentials
\cite{Miyazawa85_M,SamudralaMoult98_JMB,LuSkolnick01_Proteins}, and
empirical potentials obtained from optimization
\cite{Goldstein92_PNAS,ThomasDill96_PNAS,Bastolla01_Proteins,Dima00_PS,Micheletti01_Proteins}.

The effectiveness of potential function depends on another critically
important factor, {\it i.e.}, the representation of protein
structures.  Within this framework, we explore a new type of pairwise
contact potentials using a novel contact definition.  We introduce a
contact definition that reflects protein geometry more accurately.
These contacts are based on the computation of the alpha shape, or the
dual simplicial complex description of protein structures
\cite{EdMu94,Liang98a_Proteins}.  Here contact occurs if
atoms from non-bonded residues share a Voronoi edge, and this edge is
at least partially contained in the body of the molecule.  In another
word, atoms have to be in physical nearest neighbor contact.  In this
study, we only examine the 1-simplices, or edges in the dual
simplicial complex, which represents the pairwise contacts.  This
description is related to the work from Wodak's group, where the
contact area between atoms are calculated as the area of intersection
of the accessible atom ball around each atom and the faces of its
weighted Voronoi cell \cite{Wodak_Proteins1999}.

We develop a statistical contact propensities based on alpha edge
simplices and a combinatorial null model.  To account for the
uncertainty of estimated propensity parameter, we develop a studentized
bootstrap method for estimating their 95\% confidence intervals.  We
also examine how geometric representation of dual simplicial complex
influence the effectiveness of pairwise contact potential functions.
In addition, we aim to understand how the size of alphabet of amino
acid residues affect the effectiveness of empirical pair potentials.
This is important for protein design, where any reduction of the
alphabet size of residues will result in exponentially more efficient
sampling in the sequence space, therefore leading to more successful
design strategies \cite{Riddle97_Nature,FinkBall01_PRL}.

This work is also motivated by the need to develop potential functions
that takes advantage of recent development in computational geometry
and computational topology. The alpha shape representation of proteins
from computational geometry allow rapid calculations of metric,
topological, and combinatorial structures of proteins precisely
\cite{Liang98a_Proteins,Liang98b_Proteins,Edels98_DAM}.  These
advantages can lead to improvements in voids and binding site
detection \cite{Liang98b_Proteins,Edels98_DAM,Liang98_PS}, in hierarchical
representation of protein dynamic shapes at different resolution, and
in conformation sampling.  Recent theoretical development suggests
many intriguing applications in protein studies
\cite{ELZ00,EdZo01,EHZ01}.  These important advances are largely
unexploited currently, and we hope this work provides a useful link by
developing empirical pairwise contact potentials based on dual
simplicial complex representations of proteins.

Our approach also solves a problem that cannot be satisfactorily
addressed with current contact pair potentials.  In current
approaches, pairwise contact interactions are declared if two residues
are within a specific cut-off distance.  Potentials based on this
contact definition have achieved considerable success.  Nevertheless,
contacts by distance cut-off can potentially include many implausible
non-contacting neighbors, which have no significant physical
interaction \cite{Smith99_Proteins}.  Whether or not a pair of
residues can make physical contact depends not only on the distance
between their center positions (such as C$_\alpha$ or C$_\beta$, or
geometric centers of side chain), but also on the size and the
orientations of side-chains \cite{Smith99_Proteins}. Furthermore, two
atoms close to each other may in fact be shielded from contact by
other atoms. As emphasized in \cite{Taylor97_JMB}, these contact
pairs should not contribute to the assessment of pairwise contacts. By
occupying the intervening space, other residues can block a pair of
residues from interacting with each other.  Inclusion of these
fictitious contact interactions would be undesirable.

We organized this paper as follows. First, we describe briefly the 
dual simplicial complex representation of proteins structures.  We
then discuss the probabilistic models for developing pairwise
potentials.  A bootstrap resampling procedure is then introduced which
provide confidence intervals of estimated pairwise potential
parameters.  We then present the pairwise contact potential along with
experimental results in discriminating native structure against decoy
structures using several benchmark data sets.  We further examine how
pair potentials developed from the dual simplices compare with cut-off
contact definitions using side-chain centers.  The effects of reduced
alphabet size for amino acid residues are then described.  We conclude
with remarks and discussion.

\section{Model and Methods}

\paragraph{Alpha Contacts from Dual Simplicial Complex.}
Alpha shape has been successfully applied to study a number of
problems in proteins, including void measurement, binding site
characterization, protein packing, electrostatic calculations, and
protein hydrations
\cite{Edels95_Hawaii,Liang98a_Proteins,Liang98b_Proteins,Peters96_JMB,Liang98_PS,LiangDill01_BJ,Liang97_BJ,Liang98_BJ}.
Details of alpha shape have been described elsewhere, here we only
provide a brief description for completeness.

\begin{figure}[th]
\centerline{\epsfig{figure=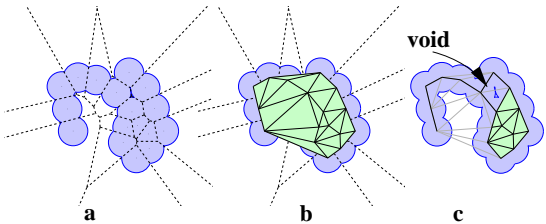,width=3in}} 
\caption{ \small \sf    
Geometric constructs of a simple 2D
molecule. a) The molecule is formed by disks of uniform size. The
dashed lines represent the Voronoi diagram, where each region contains
one atom. b) The convex hull of the atom centers, and the Delaunay
triangulation of the convex hull.  c) The alpha shape of the 2D
molecule is a subset of the Delaunay triangulation.  It is contained
within the molecule, and reflects the topological and metric
properties of the molecule.
}
\label{Fig:geometry} 
\end{figure}

To illustrate, Figure~\ref{Fig:geometry}a shows a two-dimensional
molecule formed by a collection of disks of uniform size. Each Voronoi
cell is defined by its boundaries, shown as broken lines. Every
Voronoi edge is a perpendicular bisector of the line between two atom
centers. Each Voronoi cell contains one atom, and every point inside a
Voronoi cell is closer to this atom than to any other atom. Three
connected Voronoi edges meet at a Voronoi vertex. Another geometric
construct, the Delaunay triangulation (Fig~\ref{Fig:geometry}b) is
mathematically dual to the Voronoi diagram, and can be explained by
the following procedure. For each Voronoi edge, connect the
corresponding two atom centers with a line segment, and for each
Voronoi vertex, place a triangle spanning the three atom centers of
the three Voronoi cells. Completing this for all Voronoi edges and
Voronoi vertices gives a collection of line segments and
triangles. Together with the vertices representing atom centers, they
form the ``Delaunay complex'', which is the underlying structure of
Delaunay triangulation.

Now we remove all Delaunay edges (or line segments) where
corresponding Voronoi edge of the two atoms does not intersect with
the molecular (Figure~\ref{Fig:geometry}c)). When two atoms are
spatially very close, the balls representing the two atoms intersect,
and these two atoms have non-zero, two-body volume overlap. When three
atoms are spatially very close, they intersect and have non-zero,
three-body volume overlap. We further remove all Delaunay triangles
where the corresponding Voronoi vertex of the three atoms is not
contained within the molecule. The subset of the Delaunay complex
formed by the remaining triangles, edges and vertices (atom centers)
is called the {\it dual simplicial complex}, or the {\it alpha
complex}. We are interested in identifying only contacting atoms that
are spatial nearest neighbors. These are precisely atoms with two-body
volume overlaps whose Voronoi cells intersect. By following the
mathematical dual structure, {\it i.e.}, the edges in the alpha
complex, we can accurately identify all contacting nearest neighboring
atom pairs. For convenience, we use a rather arbitrary criterion and
declare two residues are in alpha contact if there is at least one
edge connecting these two residues.

Using the alpha shape API kindly provided by Prof.\ Edelsbrunner, a
program {\sc interface} has been implemented to compute contacting
atoms based on precomputed Delaunay triangulation and alpha shape. The
Delaunay triangulation is computed using the {\sc delcx} program, and
the alpha shapes computed using the {\sc mkalf} program. Both can be
downloaded from the website at NCSA ({\tt
http://www.ncsa.uiuc.edu}). The van der Waals radii of protein atoms
are taken from \cite{Tsai99_JMB}.  We follow Singh \& Thornton and
increment the van der Waals radii by 0.5 \AA
\cite{SinghThornton92}. This increment is small and comparable with the
resolution of the structure. It enables the modeling of imprecisely
determined atomic coordinates without introducing many spurious
two-body contacts.

\paragraph{Probabilistic Model for Pairwise Alpha Contact Propensity.}
The propensity $P(i,j)$ for residue of type $i$ interacting with
residue of type $j$ is modeled as the odds ratio of the observed probability
$q(i,j)$ and the expected probability $p(i,j)$ of a pairwise alpha
contact involving both residue $i$ and $j$, :
\begin{equation}
P(i,j) = \frac{q(i,j)}{p(i,j)}
\label{equation}
\end{equation}
To compute $p(i,j)$ and $q(i,j)$, we choose a simple null model, where
the observed contacts of different proteins in the entire database are
pooled together, and the expected contact numbers are calculated.
This is the same as the
reference state of composition-independent scale discussed in
literature \cite{Skolnick00_Proteins}. We have:
\[ 
q(i,j) = \frac{a(i,j)} {\sum_{i',j'} a(i',j')} 
\] 
Here, $a (i, j)$ is the number count of atomic contacts between
residue type $i$ and residue type $j$, and $\sum_{i', j'} a (i', j')$
is the total number of all atomic contacts.  The observed probability
is then compared against the random probability $p (i, j)$ that a pair
of contacting atoms is picked from a residue of type $i$ and a residue
of type $j$, when chosen randomly and independently \cite{Adamian01_JMB}. We have:
\[ 
p(i,j) = N_{i}N_{j} \cdot (\frac{n_{i}n_{j}}{n(n-n_{i})} +
                    \frac{n_{i}n_{j}}{n(n-n_{j})}), \quad \mbox{ when}\quad \ i \neq j 
\] 
and
\[ 
p(i,j) = N_{i}N_{i-1}\cdot \frac{n_{i}n_{i}}{n(n-n_{i})}, \quad \mbox{ when }\quad
i = j 
\]
where $N_i$ is the number of interacting residues of type $i$, $n_i$
is the number of atoms residue of type $i$ has, and $n$ is the total
number of interacting atoms.

The alpha contact potential between residue $i$ and residue $j$ is
obtained from the propensity value $P(i,j)$ as $\ln P(i,j)$, and the
overall energy of a protein is calculated as:
\[
\epsilon  = -\sum_{i,j}\ln P(i,j);  
\]

\paragraph{Cys-Cys Interactions.}
In principle, only 210 parameters are necessary for 20 amino acid
residues.  However, Cys-Cys contact requires special attention.  Its
propensity value is the largest compared to the other 209 contacts,
because Cys residue tends to form disulfide bond with another Cys
residue.  Nevertheless, there are many Cys-Cys residue pairs that are
in close spatial proximity but do not form disulfide bond.  As a
result, a mis-classification of a non-disulfide Cys-Cys contact as a
disulfide bond will affect the overall score considerably, especially
for small proteins with abundant Cys residues. The problem associated
with the mis-classification of Cys-Cys contact was already pointed out
in \cite{Bastolla01_Proteins}.

\begin{table*}[!htb]
      \begin{footnotesize} \begin{center} 
\begin{tabular}{l| c c c c c
      c c c c c} \hline \hline
      &ALA&ARG&ASN&ASP&CYS&GLN&GLU&GLY&HIS&ILE\\ \hline
      ALA&\underline{2.1}&1.0&0.9&0.8&1.7&1.0&0.8&1.4&1.0&1.7\\
      ARG&0.9--1.0&\underline{0.7}&0.8&1.2&0.8&0.9&1.3&0.9&0.8&0.8\\
      ASN&0.9--1.1&0.7--0.8&\underline{1.0}&0.9&0.8&0.9&0.7&1.0&0.8&0.7\\
      ASP&0.8--0.9&1.1--1.3&0.9--1.0&\underline{0.6}&0.7&0.8&0.5&0.8&0.8&0.6\\
      CYS&1.5--1.9&0.7--0.9&0.7--0.9&0.6--0.9&$^{a}$\underline{15.2}&0.8&0.6&1.5&1.3&1.5\\
      GLN&1.0--1.1&0.8--0.9&0.8--1.0&0.7--0.8&0.7--0.9&\underline{0.9}&0.8&0.9&0.7&0.8\\
      GLU&0.8--0.9&1.2--1.4&0.6--0.7&0.4--0.5&0.5--0.7&0.7--0.8&\underline{0.5}&0.6&0.7&0.7\\
      GLY&1.3--1.5&0.9--1.0&1.0--1.1&0.8--0.9&1.3--1.7&0.8--1.0&0.6--0.7&\underline{1.5}&0.8&1.1\\
      HIS&0.9--1.1&0.7--0.9&0.7--0.8&0.8--0.9&1.1--1.5&0.6--0.8&0.6--0.7&0.7--0.9&\underline{1.0}&0.9\\
      ILE&1.5--1.9&0.7--0.8&0.7--0.8&0.6--0.6&1.4--1.7&0.7--0.9&0.6--0.7&1.0--1.2&0.8--0.9&\underline{2.1}\\
      LEU&1.5--1.6&0.8--0.9&0.7--0.8&0.6--0.7&1.4--1.7&0.9--1.0&0.7--0.7&1.0--1.2&1.0--1.1&1.9--2.0\\
      LYS&0.8--0.9&0.4--0.5&0.7--0.9&1.2--1.3&0.5--0.7&0.7--0.8&1.3--1.4&0.7--0.8&0.5--0.7&0.7--0.8\\
      MET&1.5--1.7&0.8--1.0&0.7--0.9&0.6--0.8&1.5--2.2&0.9--1.1&0.7--0.8&1.1--1.4&0.9--1.2&1.7--2.1\\
      PHE&1.2--1.4&0.8--1.0&0.7--0.9&0.5--0.6&1.5--1.9&0.8--0.9&0.6--0.6&1.0--1.2&0.9--1.1&1.5--1.7\\
      PRO&0.8--0.9&0.7--0.8&0.5--0.7&0.4--0.5&1.0--1.3&0.7--0.8&0.5--0.6&0.8--0.9&0.7--0.9&0.7--0.8\\
      SER&1.1--1.2&0.7--0.8&0.9--1.0&1.0--1.1&1.1--1.5&0.8--1.0&0.8--0.9&1.1--1.2&0.9--1.1&0.8--0.9\\
      THR&1.1--1.3&0.8--0.9&0.9--1.0&0.8--1.0&0.9--1.2&0.9--1.1&0.8--0.9&1.0--1.2&0.8--1.0&1.0--1.1\\
      TRP&1.0--1.2&1.2--1.5&0.8--1.1&0.5--0.7&1.4--2.1&0.9--1.2&0.6--0.8&1.1--1.4&1.2--1.6&1.2--1.5\\
      TYR&1.0--1.2&1.0--1.1&0.7--0.8&0.7--0.7&1.2--1.5&0.8--1.0&0.6--0.7&1.0--1.2&1.0--1.3&1.2--1.4\\
      VAL&1.5--1.7&0.7--0.8&0.7--0.8&0.5--0.6&1.4--1.7&0.7--0.8&0.6--0.7&1.1--1.2&0.8--0.9&1.8--1.9\\
      \hline
      $^{b}$$CI_{(i,i)}$&1.9--2.3&0.7--0.8&0.9--1.1&0.5--0.7&13.2--17.3&0.8--1.0&0.5--0.5&1.4--1.6&0.8--1.2&2.0--2.2\\
      \hline \hline &LEU&LYS&MET&PHE&PRO&SER&THR&TRP&TYR&VAL\\ \hline
      ALA&1.6&0.9&1.6&1.2&0.8&1.2&1.2&1.1&1.1&1.6\\
      ARG&0.9&0.5&0.9&0.9&0.7&0.8&0.8&1.3&1.1&0.7\\
      ASN&0.7&0.8&0.8&0.8&0.6&1.0&0.9&1.0&0.8&0.7\\
      ASP&0.6&1.3&0.7&0.5&0.5&1.0&0.9&0.6&0.7&0.6\\
      CYS&1.6&0.6&1.8&1.7&1.1&1.3&1.1&1.8&1.4&1.5\\
      GLN&1.0&0.8&1.0&0.8&0.8&0.9&1.0&1.0&0.9&0.8\\
      GLU&0.7&1.3&0.7&0.6&0.6&0.8&0.8&0.7&0.6&0.6\\
      GLY&1.1&0.8&1.2&1.1&0.9&1.2&1.1&1.2&1.1&1.2\\
      HIS&1.0&0.6&1.1&1.0&0.8&1.0&0.9&1.4&1.1&0.9\\
      ILE&1.9&0.7&1.9&1.6&0.7&0.9&1.0&1.3&1.3&1.8\\
      LEU&\underline{2.0}&0.7&1.8&1.7&0.7&0.9&1.0&1.6&1.3&1.7\\
      LYS&0.7--0.7&\underline{0.5}&0.7&0.7&0.5&0.8&0.7&0.9&0.9&0.7\\
      MET&1.7--1.9&0.7--0.8&\underline{2.4}&1.9&0.9&1.0&1.1&1.6&1.5&1.6\\
      PHE&1.7--1.8&0.7--0.8&1.7--2.0&\underline{2.0}&1.0&1.0&0.8&1.7&1.4&1.6\\
      PRO&0.7--0.8&0.5--0.6&0.8--1.0&0.9--1.0&\underline{0.7}&0.7&0.7&1.4&1.2&0.8\\
      SER&0.9--1.0&0.7--0.8&0.9--1.2&0.9--1.0&0.6--0.8&\underline{1.1}&1.0&1.0&0.9&0.9\\
      THR&0.9--1.0&0.6--0.7&1.0--1.2&0.8--0.9&0.6--0.7&1.0--1.1&\underline{1.1}&0.8&0.8&1.0\\
      TRP&1.5--1.7&0.8--1.0&1.4--1.9&1.5--1.8&1.3--1.5&0.9--1.1&0.8--0.9&\underline{1.8}&1.4&1.4\\
      TYR&1.3--1.4&0.9--1.0&1.4--1.6&1.3--1.5&1.1--1.3&0.8--0.9&0.8--0.9&1.3--1.5&\underline{1.2}&1.2\\
      VAL&1.7--1.8&0.6--0.7&1.5--1.7&1.5--1.6&0.8--0.9&0.8--0.9&1.0--1.1&1.3--1.5&1.1--1.3&\underline{1.8}\\
      \hline
      $CI_{(i,i)}$&1.9--2.1&0.5--0.5&2.1--2.7&1.8--2.1&0.6--0.8&1.1--1.3&1.0--1.2&1.5--2.1&1.1--1.3&1.7--1.9\\
      \hline $^{c}$$P$(CC)&\multicolumn{10}{l}{$P_{\mbox{non-disulfide-bond}} = 1.8$ with $CI=(1.7, 1.9)$; \hspace{2mm} $P_{\mbox{disulfide-bond}} = 13.3$ with $CI=(12.2, 15.3)$;}\\

      \hline \hline
\end{tabular}
\end{center}
\end{footnotesize}
\caption{\small \sf
The alpha contact potential of
pairwise interactions of amino acid residues.  The upper triangle of
the table lists the propensity values, the lower triangle lists the 95\%
confidence intervals.  The 95\% confidence intervals for the diagonal
entries are listed separately.  The potential values for the two
different types of Cys-Cys contacts are also listed.
\newline
$^a$ The alpha contact potential of Cys-Cys if all Cys-Cys
conformations are classified as one type
\newline 
$^b$ 95\% confidence interval of alpha contact potentials between
self-pair of amino acid residues.
\newline
$^c$ $P_{\mbox{non-disulfide-bond}}$ is the potential of Cys-Cys without a
disulfide bond; $P_{\mbox{disulfide-bond}}$ is the potential of Cys-Cys with
a disulfide bond.
}
\label{Tab:Potentials} 
\end{table*}

To avoid assigning the same score to the two different types of
Cys-Cys contacts, we introduce a slightly more detailed propensity
scores for Cys-Cys interactions.  Since contacts between C:O, C:N,
C:C, and O:O atoms never appear in disulfide bonds, a Cys-Cys contact
pair lacking these atomic interactions is classified as a disulfide
bond Cys-Cys contact if in addition the distance between their SG
atoms is less than 2.5 \AA.  All other Cys-Cys interactions are
classified as nonbonded Cys-Cys contact.  The propensity values
estimated for these two types of Cys-Cys contacts are listed in
Table~{\thetable{\ref{Tab:Potentials}}}.

\paragraph{Bootstrap Resampling.} 
Because the sample size of 1,045 proteins in {\sc pdbselect } is
limited, statistical modeling with approximations may be prone to errors.
It is therefore essential to assess reliability of estimated contact
potentials.  Here we apply bootstrap techniques to calculate
confidence intervals from simulated data sets \cite{Efron1993,Davison97}.  For alpha contact
potential of a specific residue pair, ({\it e.g.}, Trp-Trp), we denote
the true value of the contact potential as $\theta$.  Our
probabilistic model (Equation~\ref{equation}) can be regarded as an
estimator $T$ that gives the estimated value $t$ from the finite
amount of data for $\theta$.  Our goal is to calculate a
$95\%$  confidence interval for $\theta$.

We resample 1,045 proteins independently $R$ times from the
set of 1,045 proteins from {\sc Pdbselect }, with replacement allowed.
We have a simulated data set of $Y^*_1, .., Y^*_R$, each contains 1,045
proteins.  Some structures in the original {\sc Pdbselect } set appear
multiple times, some appear once, and some never appear.  We estimate
the pair contact value $\theta$ from each of the $R$ samples, and
obtain $t^*_1, .., t^*_R$.  For an equitailed $95\%$ confidence
interval ($ 95\%= 1-2\alpha, \alpha = 2.5\%$), we have the {\em basic
bootstrap confidence limits}:
\[
(t^*_{(R+1)(1-\alpha)}, \quad t^*_{(R+1)\alpha})
\]
The accuracy of these limits depend on $R$, and how well the
distribution of $\{t^* - t\}$ agrees with that of $T-\theta$.  Perfect
agreement occurs only when the distribution of $T - \theta$ does not
depend on any unknown variables.

To reduce possible errors due to unknown variables, we use {\it
studentized bootstrap}.  For the $r$th bootstrapped sample, we
calculate:
\[
z^*_r = \frac{t^*_r - t}{v^{* 1/2}_r}
\]
To obtain $v^*_r$, we bootstrap with replacement again $M$ times the
$r$th sample of the original bootstrap.  We have:
\[
v^*_r = \frac{1}{M-1}\sum_{m=1}^M (t^*_m - \bar{t}^*)^2
\]
where $t^*_1,...t^*_M$ are calculated from the second bootstrap
sampling.  We then use the $(R+1) \cdot \alpha$th order statistic of the
simulated values $z^*_1, ...,z^*_R$, or $z^*_{(R+1)\alpha}$ to obtain
the studentized bootstrap confidence interval for $\theta$:
\[
(t - v^{1/2}z^*_{(R+1)(1-\alpha)}, \quad
t - v^{1/2}z^*_{(R+1)\alpha})
\]

Since $M$ bootstrap samples from the $r$th sample are needed to obtain
$v^*_r$, the total number of nested bootstrap samples is $(M+1) \cdot
R$.  We chose $R = 1,000$ and $M = 50$.  Altogether, we generate
$1,001 \times 50 = 50,050$ bootstrap samples to calculate the
confidence intervals for each of the $209 + 2$ pairwise alpha contact
propensities.

\paragraph{Database Selection.}  In this study, we use {\sc Pdbselect } 
from {\tt http://www.cmbi.kun.nl /swift/pdbsel}
\cite{Hobohm_ProSci1992,Hobohm_ProSci1994}.  {\sc Pdbselect } contains
1,045 proteins selected from the Protein Data Bank (PDB). The sequence
identity between any pair of proteins in {\sc Pdbselect } is smaller
than 25\%.

\section{Results}

\paragraph{Pairwise Alpha Contact Potentials.}

The pairwise alpha contact propensities are listed in
Table~{\thetable{\ref{Tab:Potentials}}}.  These propensities are
calculated for all residue contacts that are at least 3 residues away
in primary sequence.  As expected, Cys-Cys has the highest propensities
for contact interactions.  Other residue pairs with the highest
propensities for contact interactions ($P(i,j) = 1.4-2.5$) are pairs
of hydrophobic residues ({\it e.g.}, Met-Met, Ala-Ala, Ile-Ile,
Phe-Phe, Ile-Leu, and Ile-Met). The group of residue pairs with the
second highest propensities ($P(i,j) = 1.2-1.3$) are ionizable
residues with opposite charges ({\it e.g.}, Arg-Asp, Arg-Glu, Asp-Lys,
and Gly-Lys).  Residue pairs with lowest alpha contact propensities
($P(i,j)=0.4-0.6$) are dominated by pairs of ionizable residues of the
same charge ({\it e.g.},Arg-Lys, Asp-Glu, Lys-Lys, and Glu-Glu).  The
group of residue pairs with the second lowest alpha contact
propensities ($P(i,j) = 0.5-0.7$) are between ionizable residues and
hydrophobic residues ({\it e.g.}, Asp-Phe, Asp-Ile, Asp-Leu, and
Glu-Val).  Noticeably, pairs of Pro and ionizable/polar residues also
have very low propensity for contacting interactions, probably due to
the lack of a backbone-NH for H-bonding interactions.

The confidence intervals of these propensity values given by the
studentized bootstrap procedure indicate that most of them are estimated
accurately.  Among the 209 parameters excluding Cys-Cys interaction,
the 95\% confidence intervals for 153 contact propensities are $<0.2$, a
very tight interval.  The confidence intervals of 36 contact
propensities are $<0.3$.  Contact propensities with the largest confidence
intervals around 0.6 are Trp-Trp, Met-Met, Cys-Met, and Cys-Trp.

\paragraph{Correlating and Clustering Similar Amino Acid
Residues.}

The overall behavior of pairwise contact interactions for a specific
residue type is determined by its 20 pairwise contact propensity
values.  These values represent a profile of contact interactions
specific for the residue type, and can be represented as a
20-dimensional vector $\bx$.

\begin{figure}[thb]
\centerline{\epsfig{figure=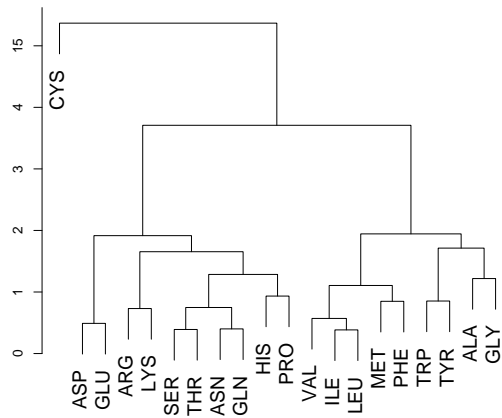,width=3in}} 
\caption{\small \sf
Grouping of the 20 types of amino
acid residues by hierarchical clustering with complete linkage of the
Euclidian distance between their 20-dimensional propensity vectors.
}
\label{Fig:clustering}
\end{figure}

We group the 20 types of amino acid residues by the Euclidean distance
between the 20 vectors \cite{Adamian01_JMB}. Fig~\ref{Fig:clustering}
shows the grouping of the 20 amino acid residues obtained by
hierarchical clustering. As an exploratory tool for data analysis,
hierarchical clustering can discover interesting and informative
grouping patterns in data \cite{Kaufman90}. In
Fig~\ref{Fig:clustering}, residues that have close contact propensity
values to the 20 residue types are grouped together.

The pattern of groupings of residues clearly reflect the physical and
geometric characteristics of the amino acid residues.  Cys residue is
different from all other residues, because of its propensity to form
disulfide bond.  The rest of the 19 residues can be broadly divided into
two well defined branches of hydrophobic and hydrophilic residues.
Among the hydrophilic residues, ionizable residues with positive
charge and negative charge are grouped into two individual small
clusters.  Hydroxyl-containing residues (Ser and Thr) and amide
residues (Asn and Gln) are also clustered into two individual small
clusters.  These residues are all capable of forming side chain
hydrogen bonds, and their clusters are neighboring with each other,
forming a larger cluster of Ser, Thr, Asn, and Gln.  Among the
hydrophobic residues, the branched residues (Val, Ile, and Leu), and
small residues (Ala and Gly) are grouped into their own clusters.
Aromatic residues Trp and Tyr also group into one cluster.  Phe has
strong hydrophobicity and is group with other strongly hydrophobic
amino acid residues, rather than clustering with the other two
aromatic residues. Pro and His are grouped together, probably because
both do not form strong favorable contacts with either hydrophilic or
hydrophobic residues.

The clustering pattern of residues by alpha contact propensity also
resembles to a certain extent the clustering pattern derived from
mutation matrix {\sc BLOSUM50} \cite{Henikoff92_PNAS} as
reported by Murphy {\it et al} \cite{Murphy&levy_ProEng2000}.  For
example, Arg with Lys, Asn with Gln, Ser with Thr, and Glu with Asp
are all clustered tightly with each other. In addition to the distinct
grouping of Cys in our clustering result, a notable difference is that
by alpha contact propensity residues Ser, Thr and Pro are grouped with
hydrophobic residues instead of hydrophilic residues.

\paragraph{Discriminating Native Structures from Decoys.}
An important method to determine the effectiveness of contact
potential functions is to evaluate its success and failure in
distinguish native protein structures from incorrectly folded decoy
structures.  We use three decoy data sets to assess alpha contact
potential.

\begin{table}[tb]
      \begin{small}
      \begin{center}
      \begin{tabular}{l c c c c}
      \hline \hline
      &Misfold&ifu&Asilomar&pdberr \\
      &       &   &        & and spga\\
      \hline
      RAPDF     &25/25&32/44&39/42&5/5\\
      CDF       &19/25&20/44&36/42&5/5\\
      GC        &25/25&21/44&32/42&4/5\\
      MJ        &25/25&21/44&34/42&5/5\\
      BT        &25/25&22/44&35/42&4/5\\
      Alpha     &25/25&24/44&37/42&4/5\\
      \hline
      \hline
      \end{tabular}
      \end{center}
      \end{small}
\vspace{2mm}
      \caption{ \small \sf
Discriminating native structure from
decoys in {\sc ProStar} data sets using alpha contact potential
(Alpha).  Results for each decoy subset are compared with those of
other potentials, including RAPDF (Residue-specific All-atom
Conditional Probability Discriminatory Function)
\cite{SamudralaMoult98_JMB}, CDF (Contact Discriminatory Function)
\cite{SamudralaMoult98_JMB}, MJ (Miyazawa \& Jernigan Potential)
\cite{Miyazawa85_M}, BT (Betancourt \& Thirumalai Potential)
\cite{Betancourt99_PS} and GC (Geometry Center Based Potential).  The
first number in each cell is the number of correctly identified native
proteins, and the second number is the total number of proteins in the
subset.  The first row lists the names of the decoy subsets.  Data for
RAPDF, and CDF are taken from Figure 1b of reference
\cite{SamudralaMoult98_JMB}.
}
      \label{Tab:Prostar}
\end{table}

{\it ProStar.}  The decoy sets in {\sc ProStar} database
\cite{SamudralaMoult98_JMB} contain several subsets.  The {\sc
misfold} subset contains conformations of 25 sequences which are
obtained by placing these sequences on structures of different folds
with the same number of residues.  The conformations of the side
chains are obtained by Monte Carlo sampling \cite{HolmSander92}.
Alpha contact potential succeeded in identifying all 25 native
structures correctly (Table~{\thetable{\ref{Tab:Prostar}}}).

For the {\sc Asilomar} subset, alpha contact potential failed to
identify 5 native structures out of 42 proteins.  For the {\sc ifu}
subset, alpha contact potential failed for 20 out of 44 native
structures.  Alpha contact potential belongs to the class of
residue-based potentials, similar to MJ, BT, CDF
\cite{SamudralaMoult98_JMB}.  As pointed out in
\cite{LuSkolnick01_Proteins}, decoys in {\sc ifu} subset are
especially challenging for residue-based potentials, because they are
conformations of short loops, and have only a small number of residue
contact interactions.  

In the subset {\sc pdberr} and {\sc sgpa}, the decoys are structures
determined by diffraction but contain serious error, or are structures
generated by molecular dynamics simulations starting from experimental
conformations. Alpha contact potential missed one out of the five
native structures.  

      \begin{table*}[htb]
      \begin{small}
      \begin{center}
      \begin{tabular}{l l l c c l}
\hline\hline
Decoy Set & Protein &  Description & $N_{res}$&$N_{decoy}$ &   cRMSD  range    \\
\hline
4state\_&1ctf&	   C-terminal domain of the ribosomal protein L7/L12 &68  &630&   2.16 - 10.16\\
reduced &1r69&	   N-terminal domain of phage 434 repressor &63  &675&   2.28 -  9.50\\
        &1sn3&      Scorpion toxin variant 3 &65 &660 &  2.50 - 10.46\\
        &2cro&      phage 434 Cro protein &65 &674   &2.05 -  9.72\\
        &3icb&      Vitamin D-dependent calcium-binding protein &75 &653 &  1.81 - 10.74\\
        &4pti&      trypsin inhibitor &58 & 687  & 2.83 - 10.79\\
\vspace{3mm}
        &4rxn&      rubredoxin &54 &677   &2.58 -  9.28\\
\hline
lattice\_&1beo&      $\beta$ -Cryptogein &98 & 2000 & 7.00 - 15.61\\  
ssfit   &1ctf&      (see above) &68 &2000&  5.45 - 12.81\\   
        &1dkt-A&    Human Cyclin Dependent Kinase Subunit, Type 1&72 & 2000 & 6.69 - 14.05\\    
        &1fca&      Ferredoxin &55 &2000&  5.14 - 11.39\\     
        &1nkl&      Nk-Lysin &78& 2000 & 5.27 - 13.64\\      
        &1pgb&      B1 immunoglobulin-binding domain of &56& 2000 & 5.81 - 12.91\\   
                    &&streptococcal protein G&&\\  
        &1trl-A&    NMR solution structure of the C-terminal fragment &62 &2000&  5.38 - 12.52\\    
                    &&255-316 of thermolysin&&\\
\vspace{3mm}
        &4icb&      Calcium-Binding Protein&76 &2000 & 4.74 - 12.92\\      
\hline
lmds    &1b0n-B&    Sini protein subunit   &39 & 497 & 2.45 - 6.03\\ 
        &1bba&      Pancreatic hormone (AVE. NMR) &36 & 500 & 2.78 - 8.91\\
        &1ctf&      (see above)            &68 & 497 & 3.59 - 12.53\\
        &1dtk&      Dendrotoxin K (NMR)     &57 & 215 & 4.32 - 12.58\\
        &1fc2-C&    Fragment B of protein A (Complexed to &43&500& 4.00 - 8.45\\
        &      &    immunoglobin Fc)&&&\\
        &1igd&      3rd IgG-binding domain from streptococcal protein G&61 &500 & 3.11 - 12.56\\ 
        &1shf-A&    Fyn Proto-Oncogene Tyrosine &59 &437& 4.39 - 12.35\\
        &      &    Kinase subunit (SH3 domain)&&&\\
        &2cro&      (see above) & 65 &500 &3.87 - 13.48\\
        &2ovo&      3rd domain of silver pheasant ovomucoid &56 &347 & 4.38 - 13.38\\
        &4pti&      (see above) &58 &343 &4.94 - 13.18\\
     \hline\hline
      \end{tabular}
      \end{center}
      \end{small}
      \vspace{4mm}
      \caption{\small \sf
 Description of proteins in the {\sc
4-state-reduced} decoy set, {\sc Lattice-ssfit} decoy set and {\sc
Lmds} decoy set. The number of decoy structures and the cRMSD ranges
are listed.}
      \label{Tab:rmsdrange}
\end{table*}

{\it Park \& Levitt Set.}  This decoy test set contains native and
near-native conformations of seven sequences, along with about 650
misfolded structures for each sequence.  The positions of $C_\alpha$
of these decoys were generated by exhaustively enumerating ten
selectively chosen residues in each protein using a 4-state
off-lattice model. All other residues were assigned the phi/psi value
based on the best fit of a 4-state model to the native
chain. Conformations in the decoy sets all have low score by a variety
of scoring functions, and have low RMSD to the native structure
(Table~{\thetable{\ref{Tab:rmsdrange}}}) \cite{ParkLevitt96_JMB}.

\begin{table*}[htb] 
      \begin{small} 
      \begin{center} 
      \begin{tabular}{llccccccccccccc}
      \hline \hline
      &&\multicolumn{5}{c}{$H_\alpha$}&\multicolumn{3}{c}{$H_{gc}$}&\multicolumn{3}{c}{$H_{MJ}$}&\multicolumn{2}{c}{$H_{BT}$} \\
      \cline{3-15} 
      Decoy set & PDB & $^a$Rank &$^bZ$&$^c\overline{r}$&$^df/1,000$&& Rank &$Z$&& Rank &$Z$&& Rank &$Z$\\
      \hline 
      4state\_&1ctf&     1& 3.08&1.0&1,000& &1&3.42& &1&3.73& &1&3.86\\       
      reduced &1r69&     1& 3.33&1.0&1,000& &8&2.34& &1&4.11& &1&4.47\\
              &1sn3&     1& 3.10&1.0&1,000& &8&2.49& &2&3.17& &6&2.97\\
              &2cro&     1& 3.00&1.0&1,000& &2&2.91& &1&4.29& &1&3.92\\ 
              &3icb&     5& 2.19&3.4&52  & &10&2.14& &2&2.80& &1&2.83\\
              &4pti&     1& 2.30&1.0&1,000& &11&2.28& &3&3.16& &5&2.65\\
              &4rxn&     51&1.22&43.0&0  & &1&2.75& &1&3.09& &1&3.01\\

\hline
      lattice\_&1beo&    1& 4.74&1.1&923 & &2&3.69& &1&4.74& &1&7.29\\     
      ssfit   &1ctf&     1& 4.62&1.0&1,000& &1&5.09& &1&5.35& &1&6.99\\
              &1dkt-A&   1& 4.33&1.0&1,000& &15&2.38& &32&2.41&&5&3.49\\
              &1fca&     40&2.01&32.0&0  & &254&1.18& &5&3.40& &2&3.92`q\\
              &1nkl&     1& 5.21&1.0&1,000 & &1&7.20& &1&5.09& &1&7.28 \\
              &1pgb&     1& 3.31&1.0&964& &32&2.18& &3&3.78& &2&3.82\\
              &1trl-A&   5& 3.35&6.2&0   & &504&0.63& &4&2.91& &2&3.82\\
              &4icb&     1& 4.59&1.0&1,000& &1&4.11& &1&3.67& &1&5.07\\
\hline
      lmds    &1b0n-B&   2& 3.13&1.5&525 & &99&0.85& &1&2.65& &2&2.50\\ 
              &1bba&     217&0.03&340.8&0& &441&-1.11& &364&-0.64& &234&0.04\\
              &1ctf&     1&3.12 &1.0&1,000& &74&1.09& &1&3.86& &1&3.15\\ 
              &1dtk&     2&2.13 &2.2&234 & &173&-0.92& &13&1.71& &122&-0.08\\ 
              &1fc2-C&   500&-3.68&500&0 & &480&-1.63& &501&-6.24& &501&-5.11\\   
              &1igd&     9&2.43&14.0&0  & &138&0.61& &1&3.25& &1&3.76\\
              &1shf-A&   17&1.46&8.2&0  & &322&-0.57& &11&1.30& &16&1.06\\
              &2cro&     1 &4.36&1.0&999& &159&0.44& &1 &5.07& &1&4.01\\
              &2ovo&     3 &3.07&5.2&29 & &326&-1.34& &2&3.25& &31&1.29\\
              &4pti&     9&2.23&7.0&0   & &242&-0.49& &4&2.53& &117&0.42\\
      \hline \hline
      \end{tabular}
      \end{center}
      \end{small}
\vspace*{3mm}
\caption{
\small \sf Discriminating native structures using
alpha contact potential $H_\alpha$ and potential by cut-off distance
between geometric centers of side chains $H_{gc}$.  The
4-state-reduced,  {\sc Lattice-ssfit}, and {\sc Lmds} decoy sets are tested.
\newline
$^a$ Rank of native structures.
\newline
$^b$ $z = \overline{E} -
E_{native}/{\sigma}$; $\overline{E}$ and $\sigma$ are the mean and
standard deviation of the energy values of conformations, respectively.
\newline
$^c$ Average ranking of native structures in energy evaluated using
1,000 bootstrapped potential values;
\newline $^d$ The number of times of a
native structure is ranked to have the lowest energy.
}
\label{Tab:Result} 
\end{table*}

\begin{figure}[t]
\centerline{\epsfig{figure=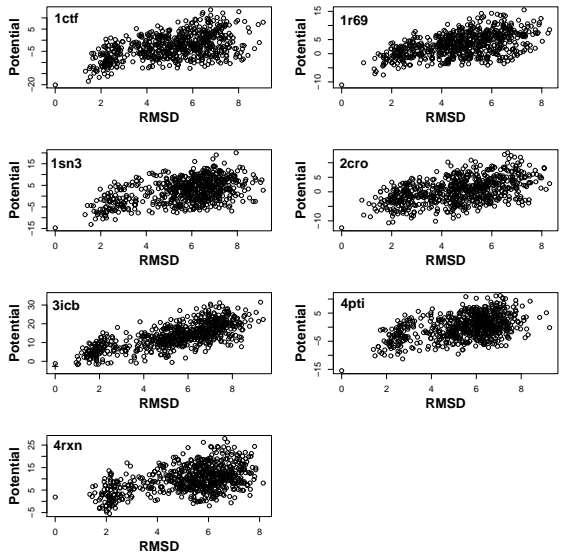,width=3in}} 
\caption{\small \sf
Energy evaluated by alpha contact potential plotted against the RMSD
to native structures for conformations in Park\& Levitt Decoy Set.
The alphabet of residues has 20 types of amino acids. For vitamin
D-dependent calcium-binding protein ({\tt 3icb}, a structure with
better resolution ({\tt 4icb}) has the lowest energy (denoted by
``+'').
} 
\label{Fig:PL} 
\end{figure}

The results of discrimination test are listed in
Table~{\thetable{\ref{Tab:Result}}} and plotted in Fig~\ref{Fig:PL}.
For five of the seven proteins, the native structures have lowest energy by
alpha contact potential. For protein {\tt 3icb} and {\tt 4rxn}, the
native structures have the 5th and 51st lowest energy values,
respectively.  For all proteins, decoys with the lowest energy are
within 2.5 \AA\ RMSD to the native structure.

The protein {\tt 3icb} is a vitamin D-dependent calcium-binding
protein. Although the energy of the native structure ranks as the 5th
lowest energy, the RMSDs of the first four lowest energy decoys are all
within 2.0 \AA\ RMSD to the native structure. For a higher resolution
structure ({\tt 4icb} at 1.60 \AA) of this protein, the energy of {\tt
4icb} by alpha contact potential is lower than any of the decoys.  It
is possible that this mis-classification might be due to the lower
resolution of structure of {\tt 3icb}.

Rubredoxin ({\tt 4rxn}) is an iron-sulfur protein.  A Fe(III) ion is
covalently bound in this structure with four Cys sulfur atoms,
preventing these four Cys from forming two possible disulfide
bonds. Because the protein description and the contact potential do
not contain any information about the important covalent bonds of Cys
with Fe-S cluster, it is reasonable to expect that native structure
will not be of lowest energy if these important covalent bond
interactions are unaccounted for.  It is possible that the structure
of rubredoxin might be different without the Fe-S cluster.  The decoys
at the lowest energy states form one or two fictitious disulfide
bridges, and all are near-native structures with RMSD around 2 \AA\ to
the native structure. MJ and BT potential work better on {\tt 4rxn}
because they classify the four Cys-Cys contacts as disulfide bonds.

{\it Lattice\_ssfit Set.}  The {\sc lattice\_ssfit} set contains
conformations for eight small proteins generated by {\it ab initio}
protein structure prediction methods
\cite{Samudrala99_Abinitio,XiaLevitt00_JCP}.  The conformational space
of a sequence was exhaustively enumerated on a tetrahedral lattice.  A
lattice-based scoring function was used to select the 10,000 best
scoring conformations. Secondary structures were fitted to these
conformations using a 4-state model \cite{ParkLevitt96_JMB}.  The
10,000 conformations were further scored using a combination of an
all-atom scoring function \cite{SamudralaMoult98_JMB}, a hydrophobic
compactness function, and a one-point per residue scoring function
\cite{ParkHuangLevitt97_JMB}. The 2,000 best scoring conformations for
each protein are selected as decoys for this data set.
      
The result (Table~{\thetable{\ref{Tab:Result}}}) shows that for six
out of eight proteins, no decoy structures score better than the
native structure. The exceptions are {\tt 1fca} and {\tt 1trl}.
Similar to {\tt 4rxn} in the Park \& Levitt decoy set, ferrodoxin {\tt
1fca} contains a Fe-S cluster. Its four Cys residues form four
covalent bond with the four Fe(III) irons, instead of two disulfide
bridges. These critical contacts again are unaccounted for in the
alpha contact potential, and therefore the native structure of this
protein was not identified successfully.  {\tt 1trlA} is a NMR
solution structure of the C-terminal fragment (255-316) of
thermolysin.  NMR structures are far more difficult to recognize, as
discussed in detail in \cite{Bastolla01_Proteins}. They are usually
represented as an ensemble of conformations. The contact energies of
conformations in the ensemble can be substantially different. It is
conceivable that an energy function valid for crystal structures
cannot reliably recognize native NMR structures
\cite{Bastolla01_Proteins}.  In addition, the structure {\tt 1trlA}
occurs in a dimeric state in the original PDB file. There is
substantial interaction between the two chains.  Because of this, it
is unclear whether single subunit of {\tt 1trlA} in monomeric state
would retain the same conformation.

The decoy structures in this data set are generated by {\it ab
initio\/} methods. None of them are near-native, and all have cRMSD to
the native structure greater than 4.7 \AA\
(Table~{\thetable{\ref{Tab:rmsdrange}}}).  When decoys are so
different from the native conformation, energy evaluated using alpha
contact potential shows little correlation with the RMSD.  The lowest
energy decoys in this data set all have large RMSD, similar to results
reported in \cite{SamudralaLevitt00_PS} (data not shown).

{\it Lmds Set.}  The local minima decoy set ({\sc lmds}) contains
decoys that were derived from the experimentally obtained secondary
structures of ten small proteins that belong to diverse structural
classes. Each decoy is a local minimum of a ``hand made'' energy
function
\cite{Levitt&Lifson_JMB1969,Levitt_JMB1974,Levitt_JMB1983,Levitt&Daggett_CompPhysComm1995}.
Ten thousand initial conformations were generated for each protein by
randomizing the torsion angles of the loop regions
\cite{Fletcher_ComputJ1970}. The adjacent local minima were found by
truncated Newton-Raphson minimization in torsion space. Each protein
is represented in the decoy set by its 500 lowest energy local minima.

The alpha contact energy function works fairly well in the recognition
of {\tt 1b0n-B, 1ctf, 1dtk, 2cro} and {\tt 2ovo}. {\tt 1dtk}
(Dendrotoxin K) contains three disulfide bonds in its native
structure. However, in most of its 215 decoys, six Cys residues are
spatially arranged together and form on average seven Cys-Cys
contacts.  For some decoys, up to 15 Cys-Cys contacts by distance
cut-off can be found.  The ability of discriminating disulfide bonded
{\it versus} non-disulfide bonded Cys-Cys contacts probably makes the
alpha contact potential discriminate better than the MJ and the BT
potentials. The native structure of {\tt 1igd} (IgG-binding domain
from streptococcal protein G) ranks first by the MJ and the BT
potential, but ranks 9th by alpha contact potential. The recognition
of {\tt 1bba} and {\tt 1fc2-C} in this set failed for all
residue-based contact potentials. {\tt 1bba} is an atypical structure
of a small protein determined by NMR, which forms a helix with random
coil. {\tt 1fc2-C} is a fragment of protein complexed with an
immunoglobulin molecule. It possible that this protein may not
maintain the same conformation without the complexed immunoglobulin.

By the criterion of the ranking of native protein, with the exception
of the ion-sulfur proteins {\tt 4rxn} and {\tt 1fca}, the overall
results shown in Table~{\thetable{\ref{Tab:Prostar}}} and
Table~{\thetable{\ref{Tab:Result}}} indicates that the performance of
alpha contact potential in discriminating native protein from decoys
is better than that of MJ and BT potentials for the {\sc misfold, ifu,
asilomar, 4state\_reduced} sets and the {\sc lattice\_ssfit} set, and
has comparable results for the {\sc lmds} set, the {\sc pdberr} set,
and the {\sc spga} set.

\section{Discussion}

\paragraph{Contact Definition.}
The alpha contacts introduced in this work are different from contacts
by cut-off distances.  Atoms in alpha contacts are all within a
distance that depends on the identities of the two atoms. In this
study, this distance is equal to the sum of the van der Waals radii of
the two atoms, plus $2\times 0.5$ \AA.  Unlike contacts by distance
cut-off, this distance is not a single fixed constant but depends on
the atom types. Another important distinction of alpha contact is that
only a subset of atoms satisfying the distance criterion will be
counted as physical nearest neighbors, because we have an additional
criterion: contacting atoms must have intersecting Voronoi cells.
Alpha contacts represents the geometry more accurately and can capture
contact interactions due to side chain size and orientation
\cite{Taylor97_JMB}.  In addition, no fictitious contacts will be
introduced between two atoms when there is a third intervening atom
\cite{Smith99_Proteins}.  Perhaps this is the reason why alpha contact
potential is sensitive to the presence of Fe-S clusters and other
hetero atoms, which can be potentially exploited for determining
whether a protein structure should contain hetero atoms.

For the 1,045 proteins in the {\sc pdbselect} dataset, we compare
contacts identified by distance cut-off with the threshold of two van
der Waals atom radii plus $2\times 0.5$ \AA\, and contacts identified
by the alpha shape.  We found that about 30 -- 50 \% of atom contacts
detected by distance cutoffs are blocked by a third atom and hence do
not have physical interactions.  As a result, 3 -- 6\% residue
contacts detected by distance cutoffs do not interact physically.
Inclusion of these fictitious contact is problematic, especially in
developing all-atom contact potentials, as well as in future studies
when higher order interactions in the form of three or four body
contacts are incorporated.

\paragraph{Evaluating Discrimination of Alpha Contact Propensities by Bootstrap.}
How robust is the results of decoy discrimination to the specific
values of alpha contact potentials and the specific choices of the
structures in the database?  We further make use of the bootstrap
resampling technique to evaluate the reliability of the discrimination
results.  As discussed earlier, we resampled 1,045 proteins in {\sc
Pdbselect} independently $R = 1,000$ times with replacement allowed,
and obtain 1,000 contact propensity matrices.  Each is then used to
discriminate the decoys in Table III.  We use two parameters:
$\overline{r}$, the average ranking of the native structure, and $f$,
the times a native structure ranked as the structure with the lowest
energy.  Table IV shows that for many decoy sets ({\it e.g.}, {\tt
1ctf}, {\tt 1r69}, {\tt 1sn3}, {\tt 2cro}, and {\tt 4pti} in the {\sc
4state\_reduced} set), the native structure always ranks first in the
1,000 bootstrapped energy evaluations.  The performance using 1,000
different sets of ``bootstrapped'' potential values validates the
robustness of the method deriving the alpha contact potential and the
informativeness of the underlying protein structure database.

\paragraph{Comparison with Contact by Geometric Centers.}

Contact definition by distance cut-off is widely used in the
development of many potential functions.  Here we compare potentials
obtained by alpha contact, denoted as $H_\alpha$, and by contact
defined by cut-off distance between geometric centers of side chains,
denoted as $H_{gc}$.  Following reference \cite{TobiElber00_Proteins}, two
residues are declared to be in contact if the distance $d$ between the
geometric centers of their side chains is: 2\AA $<d<$ 6.5\AA.
Geometric center based contact potential $H_{gc}$ are developed using
the same {\sc Pdbselect} data, with the same null model as that of
alpha contact potential, and similarly counting only contact between
residues that are three or more residues apart.  Therefore, any
difference between $H_\alpha$ and $H_{gc}$ is solely due to different
geometric representation.

\begin{figure}[t]
\centerline{\epsfig{figure=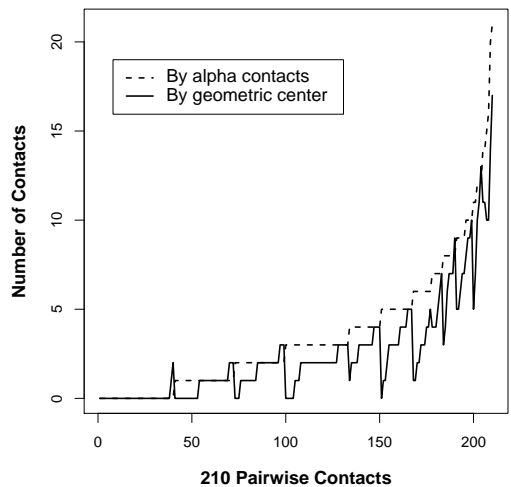,width=2.8in}}
\caption{\sf \footnotesize 
 Difference in contact histograms
between the two contact definitions of alpha contact and contact by
geometric centers for protein structure {\tt 1abe}.  Data along
$x$-axis is arranged in ascending order by the frequency of contact
pairs in the alpha shape contact model.  There are frequently
significant less contacts when contact defined by distance between
geometric centers is used.
} \label{Fig:diffContact}
\end{figure}

The $\log$ values of the parameters of $H_\alpha$ and $H_{gc}$ 
have an overall correlation coefficient of $\rho =
0.77$. However, the contact maps of individual proteins by these two
different contact definition are often substantially different.  As an
example, alpha shape dual simplicial complex gives significantly more
contacts than cut-off distance by geometric centers of side chain for
protein {\tt 1abe} (Fig~\ref{Fig:diffContact}).

\begin{figure}[t]
\centerline{\epsfig{figure=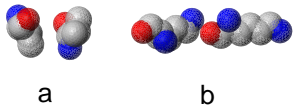,width=3in}}
\caption{\sf \footnotesize 
Alpha contact maps provide accurate
geometric descriptions.  a) Two Val residues are nearly parallel to
each other. Their geometric centers are close enough, but no physical
contacts occur between them. b) Two Lys are oriented linearly in
opposite direction. Their geometric centers are far away from each
other, but a hydrogen bond forms between O from one Lys and N from the
other Lys.  Alpha contact definition correctly identify a) as ``in
contact'' and b) as not ``in contact''.  Contact model by distance
cut-off of geometric centers of side chains give different contact
assignment in both cases.
} \label{Fig:contact}
\end{figure}

Fig~\ref{Fig:contact} illustrates why such a discrepancy exists
between these two contact models (see Figure 5 legend).  The strong
correlation between $H_\alpha$ and $H_{gc}$ is deceptive and this is
reflected in another aspect.  Although the correlation coefficient is
high, the pairwise contact potentials may have very different values
for $H_\alpha$ and $H_{gc}$.  $H_{gc}$ categorically gives much higher
propensity values for interactions between small residues.  For
example, $H_{gc}$ for Gly-Gly and Ala-Gly are 4.55 and 3.04,
respectively, but $H_\alpha$ are only 1.48 and 1.39,
respectively. Gly-Pro interaction is strongly favorable by $H_{gc}$
($P(Gly, Pro) = 1.84$), but is unfavorable by $H_\alpha$ (0.87).  On
the other hand, $H_{gc}$ gives much lower propensity values for
interactions between large residues.  For example, $H_{gc}$ for
Trp-Trp and Phe-Trp are 0.39 and 0.56, respectively, but $H_\alpha$
are 1.75 and 1.65, respectively. In addition, many pair contact
interactions between Trp and another residue with large side chains
(such as Arg, Tyr, Phe, His, Leu, Ile, and Met) are unfavorable
($<1.0$) by $H_{gc}$, but are favorable by $H_\alpha$.

These differences lead to different discrimination in identifying
native and near-native protein structures from decoy structures.
Because it is impossible to define refined potential for Cys-Cys
contacts in $H_{gc}$ model as in alpha contact model, we exclude
proteins containing Cys-Cys contacts to avoid complication for
comparison.  Table~{\thetable{\ref{Tab:Result}}} shows that $H_{gc}$
can only recognize two native structures out of nine proteins in the
union set of the 4-state decoy set and the lattice-ssfit decoy set.
In addition, the $z$ scores for native structures are generally higher
when using $H_\alpha$ than $H_{gc}$.  For the 4-state decoy set, the
correlation coefficient $\rho$ by $H_\alpha$ is also higher.  These
results indicate that geometric description of protein structures are
important, and contact model by alpha shape are more accurate with
more discriminative information for identifying native-like
structures.  However, the cut-off distance approach may be more
convenient to implement, and it is possible to gain further
improvement by setting the values of the cut-off threshold as a
variable depending on the type of contacting residue pairs.

\paragraph{Comparison with Other Potentials.}

\begin{table*}[htb]
      \begin{small}
      \begin{center}
      \begin{tabular}{l |c c c c c c}
      \hline
      \hline
            & Alpha   & MJ      & BT      & GC      &SK    &TD\\
      \hline
      Alpha &1/0      &0.66/1.45&0.80/0.28&0.77/0.41&0.61/0.51&0.66/0.39 \\
      MJ    &0.66/1.45&1/0      &0.66/1.43&0.37/1.60&0.73/1.29&0.67/1.35\\
      BT    &0.80/0.25&0.66/1.43&1/0      &0.49/0.56&0.76/0.41&0.63/0.37\\
      GC    &0.77/0.41&0.37/1.60&0.49/0.56&1/0      &0.15/0.81&0.43/0.63\\
      SK    &0.61/0.55&0.76/1.29&0.82/0.41&0.15/0.81&1/0&0.64/0.52\\
      TD    &0.66/0.39&0.67/1.35&0.63/0.37&0.43/0.63&0.64/0.52&1/0\\
      \hline \hline
      \end{tabular}
      \end{center}
      \end{small}
\vspace*{2mm}
      \caption{
\small \sf
The correlation coefficients and dispersions
\cite{Betancourt99_PS} between each pair of potentials. The first number of each cell is the correlation coefficient between each pair of potentials. The second number is the dispersion between each pair of potentials.}
      \label{Tab:ComparingPotentials}
      \end{table*}

We further compare alpha contact potential with several previously
developed residue contact potentials, including MJ
\cite{Miyazawa96_JMB}, BT \cite{Betancourt99_PS}, SK
\cite{Skolnick00_Proteins}, and TD \cite{ThomasDill96_PNAS}
potentials.  We take the values of MJ potential from Table 3 of
reference \cite{Miyazawa96_JMB}.  Following the author's
recommendation, the average hydrophobicity ({\it $e_{rr}$} = -2.55)
was subtracted from the potential, and a pair of residues is declared
to be in contact if the geometric centers of their side chain is
within an interval of 2.0 \AA\ to 6.5 \AA.  The values of BT potential
is taken from \cite{Betancourt99_PS} and was obtained by rescaling MJ
potentials with Thr as the reference solvent \cite{Betancourt99_PS}.
The correlation coefficients $\rho$ and dispersions
\cite{Betancourt99_PS} between each pair of potentials are shown in
table Table~{\thetable{\ref{Tab:ComparingPotentials}}}.  The
correlation coefficients $\rho$ for alpha contact potential $\log
P(i,j)$ and these residue contact potentials are: $\rho = 0.66, 0.80,
0.61$, and $0.66$ for MJ, BT, SK, and TD potentials, respectively.  The
dispersion as defined in \cite{Betancourt99_PS} (p.363, Formula 4) are
$1.45, 0.28, 0.51$, and $0.39$, respectively.  Because the contact map
obtained by alpha edges can be substantially different from other
contact definition (Fig~\ref{Fig:diffContact}), the absolute value of
energy by alpha contact potential and by other potentials for the same
structure can be substantially different.

\paragraph{Long Range Interactions.}
Interactions between residues with large sequence distance $d$ are
relatively rare.  We found that they occur more likely in the interior
of a protein than on the surface (data not shown).  Identifying such
interactions are of particular interest, because these interactions
result in significant reduction of conformational entropy.  Prediction
of protein structures seem to be most difficult for those with large
contact order \cite{Plaxco98_JMB}, namely, those with significant
amount of interactions between residues with large sequence distances.

The bootstrap procedure introduced here provides a method to reliably
identify the contact pairs of long sequence separation whose
propensity values can be confidently assessed (see Supplementary
Online Material for tables of alpha contact potential for $d\ge30$).
Among all possible 210 pairwise interactions, 9 contact pairs with
high propensity (lower value of 95\% confidence interval $>1.5$) can
be reliably assessed for $d>30$. In addition to Cys-Cys, they include
hydrophobic-hydrophobic interactions (Gly-Gly, Met-Met, Ile-Ile,
Phe-Phe, Val-Val, Met-Phe), salt-bridge interactions (Arg-Asp,
Asp-Lys), and Pro-Trp.

Some long range interactions are clearly associated with specific
secondary structures.  After correction for prior probability to be in
a particular secondary structure, we found that Met-Met contact has a
high propensity to occur between two helices ({\it h}) or two beta-strands
({\it s}) ($P(Met, Met)_{hh} = 2.4, P(Met, Met)_{ss} = 2.3$), and a low
propensity to occur between either a helix and a coil ({\it c})
($P(Met,Met)_{hc} = 0.65$), or a strand and a coil ($P(Met, Met)_{sc} =
0.56$). Similarly, because Gly is a helix breaker, long-range Gly-Gly
contact has a high propensity to occur between two
coils ($P(Gly,Gly)_{cc} = 3.2$), and  low propensity to occur between
two helices ($P(Gly,Gly)_{hh} = 0.53$).

\paragraph{Reduced Alphabet for Amino Acid Residues.} 

The clustering of pairwise alpha contact potentials shown in
Figure~\ref{Fig:clustering} suggests that many residues behave
similarly in contact interactions.  This points to possible degeneracy
of the amino acid alphabet \cite{Riddle97_Nature,FinkBall01_PRL}.
Reduced alphabet is important because a smaller size of alphabet will
lead to exponentially more efficient sampling methods in sequence
design and protein engineering
\cite{YueDill92_PNAS,ShakhGutin93_PNAS,Deutsch96_PRL,Li96_Science,Melin99_JCP}.
Many experimental and computational studies already suggested that a
minimum number of amino acid residue types far less than 20 may be
adequate for protein folding
\cite{Sander_JME1979,Heinz_PNAS1992,Henikoff92_PNAS,Riddle&Baker_NatStructBiol1997}.
Wang and Wang examined different ways to reduced MJ interaction matrix
and concluded that by minimizing mismatches, a reduced alphabet of
just five amino acid residue types can be used to construct sequences
with good foldability and kinetic accessibility
\cite{Wang&Wang_NatStructBiol1999}.  The reduced 5 alphabet set
coincide with the same alphabet set reported in the work of Riddle
{\it et al}, where fully functional constructs for a small 57-residue
beta-barrel protein can be experimentally obtained when residues in 38
out of 40 selected amino residues are drawn from the alphabet set of
I, K, E, A, and G.  Murphy {\it et al} further examined reduced
alphabets based on BLOSUM50 substitution matrix
\cite{Murphy&levy_ProEng2000}.  When using a variety of reduced
alphabets with size ranging from 10 to 20, they found that there is
little loss of the information necessary to select structural homologs
in a database of representative protein sequences using dynamic
programming based global alignment.

We continue investigation in this direction and study the capability
of various reduced alphabet sets in discriminating native proteins
from decoys.  Fig~\ref{Fig:clustering} provides a natural way of
reducing the residue alphabet set that is similar to the approach used
by Murphy {\it et al} \cite{Murphy&levy_ProEng2000}.  By placing a
horizontal line at different vertical heights, we can obtain a reduced
residue alphabet that is determined by the heights of the branching
points in the dendrogram from the hierarchical clustering.  For
example, we can have an alphabet of 7 residue types at height about
1.5: $\cal A$ = \{D,E\}, $\cal B$ = \{R,K\}, $\cal C$= \{S,T, N, Q, H,
P\}, $\cal D$ = \{V, I, L, M, F, W\}, $\cal E$ = \{W, Y\}, $\cal F$ =
\{A, G\}, and $\cal G$ = \{C\}.  An alphabet of two residue types
would take Cys as a residue type, and everything else grouped into
another residue type.  An alphabet of three residue types would have
Cys, polar residues, and hydrophobic residues.

\begin{figure}[tbh]
\centerline{\epsfig{figure=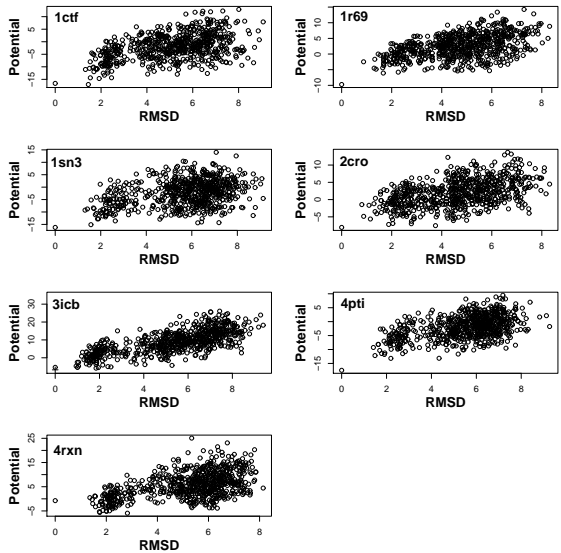,width=3in}} 
\caption{\small \sf
Energy evaluated by alpha contact
potential plotted against the RMSD to native structures for
conformations in Park\& Levitt Decoy Set.  The alphabet of residues
has 9 types of amino acids.  The discrimination is similar to that shown in
Fig~\ref{Fig:PL}. {\tt 4icb} is denoted by ``+'' and has the lowest energy.
}
\label{Fig:9set} 
\end{figure}

Does reduced alphabet still capture the basic information of protein
contact interactions?  We use Equation~\ref{equation} to estimate
pairwise alpha contact potentials for reduced alphabet size of 2, 3,
5, 7, 8, 9, 10, 11, 15, and 20, and test their effectiveness in
selecting native structure from decoys in the Park and Levitt data
set.  Fig~\ref{Fig:9set} shows the result when an alphabet set of 9
residue types, plus the three types of Cys-Cys contacts are used.
Remarkably, the discrimination of native conformation from decoys is
almost as good as when there are 20 different residue types.

\begin{figure}[hbt]
\centerline{\epsfig{figure=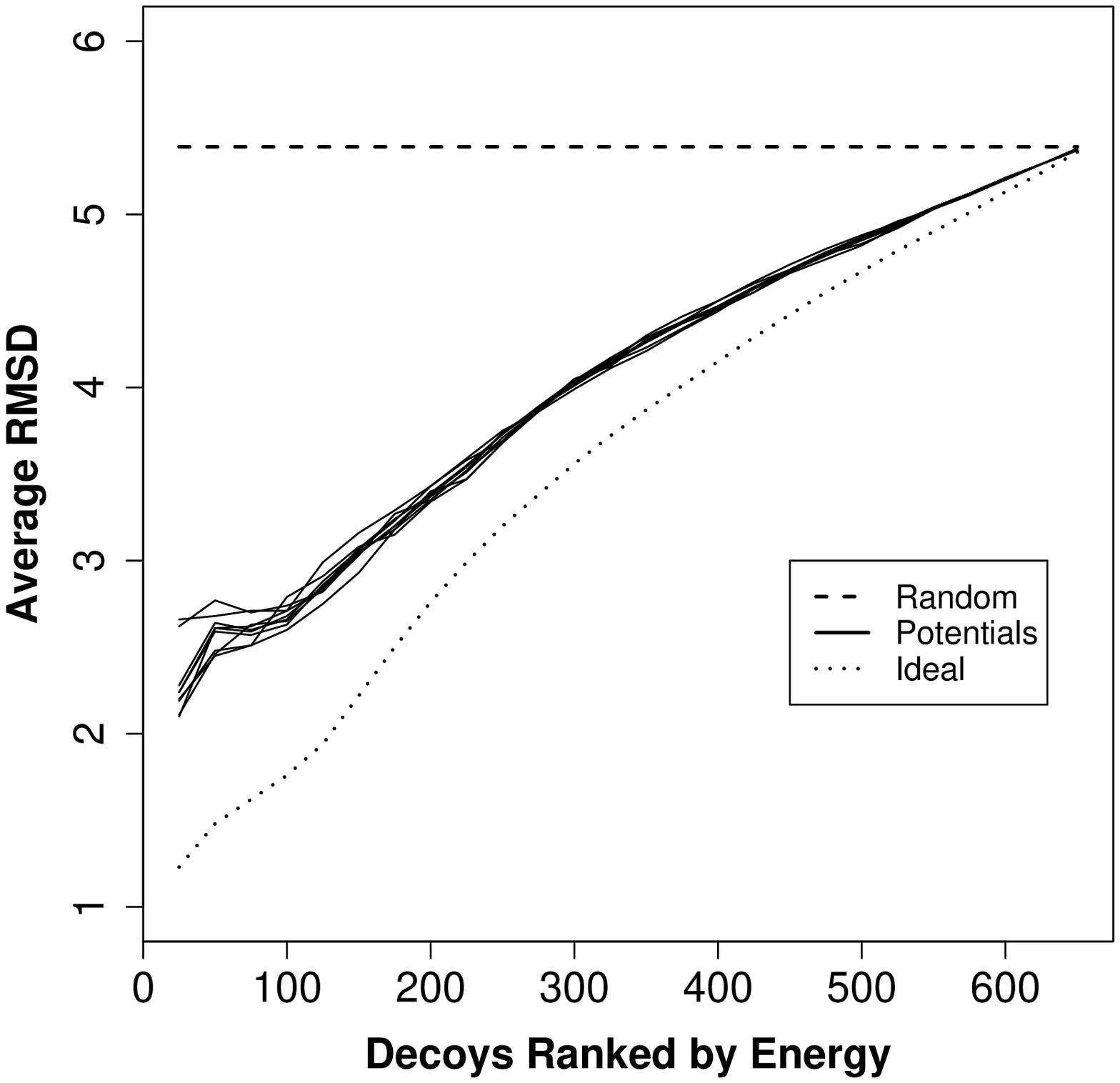,width=2.8in}} 
\caption{\small \sf
Discriminating native structure
of {\tt 3icb} from decoys using different alphabets.  For each
alphabet, we rank the decoys by their energy.  The average RMSD to the
native structure is then calculated for the top $n$ decoys at
different $n$ values.  Top dashed line represents the average RMSD if the
decoys are chosen randomly; Dotted line represents the ideal case that
$n$ decoys with smallest RMSD are chosen by an ideal function; The other
lines represent the average RMSD of decoys chosen by potentials of an alphabet with 2-20 amino acid types.
}
\label{Fig:alphabet.3icb}
\end{figure}

\begin{table*}[htb]
      \begin{small}
      \begin{center}
      \begin{tabular}{l |c c c c c c c}
      \hline
      \hline
      $^*|\Sigma|$ &1ctf&1r69&1sn3&2cro&3icb&4pti&4rxn\\ \hline
      2 &0.31& 0.35& 0.33& -0.15& 0.13 & 0.40 & 0.17\\
      3 &0.39& 0.44& 0.48& 0.34 & 0.73 & 0.44 & 0.39\\
      5 &0.41& 0.47& 0.47& 0.38 & 0.71 & 0.47 & 0.51\\
      7 &0.45& 0.44& 0.48& 0.46 & 0.71 & 0.49 & 0.48\\
      8 &0.54& 0.46& 0.42& 0.48 & 0.68 & 0.52 & 0.50\\
      9 &0.47& 0.48& 0.47& 0.47 & 0.72 & 0.47 & 0.49\\
      10&0.47& 0.49& 0.47& 0.48 & 0.71 & 0.48 & 0.51\\
      11&0.47& 0.50& 0.49& 0.47 & 0.72 & 0.49 & 0.51\\
      15&0.49& 0.50& 0.49& 0.50 & 0.71 & 0.47 & 0.49\\
      20&0.49& 0.52& 0.49& 0.50 & 0.74 & 0.46 & 0.53\\
      \hline \hline
      \end{tabular}
      \end{center}
      \end{small}
      \caption{\small \sf
 Correlation coefficients $\rho$
between energy evaluated using alpha contact potential and RMSD to the
native structures for decoys in the Park and Levitt Set.  Alpha
contact potentials with alphabet containing different number of
residue classes are used.
\newline
$^*$$|\Sigma|$ denotes the number of residue classes in the alphabet $\Sigma$ as obtained from Fig~\ref{Fig:clustering}.
}
      \label{Tab:ReducedCor}
      \end{table*}

Figure~\ref{Fig:alphabet.3icb} shows the detailed results of
discriminating native structure of {\tt 3icb} from decoys using
potentials derived from different alphabet sets.  The average RMSDs of
$n$ decoys with lowest energy by potentials of different alphabets are
calculated. Smaller average RMSD of these lowest energy decoys
indicates that a large fraction of them are near-native structures.
This would suggest good discrimination.  The top $n = 100$ decoys of
lowest energy are all found to have very similar average RMSD to the
native structure, regardless the size of the alphabet.  This suggests
that alphabet with just a few residues have respectable results in
decoy discrimination.  Table~{\thetable{\ref{Tab:ReducedCor}}} further
shows the correlation coefficient of energy and RMSD for all proteins
in the Park and Levitt data set when using different alphabets.  The
results indicate that an alphabet of 7 residues would have very
similar performance as an alphabet with 20 residues in discriminating
decoys.  Our results extended earlier work where subjectively defined
alphabet sets were used to extract contact residue potentials by an
iterative optimization method \cite{ThomasDill96_PNAS}. It was found
that potential derived using such reduced alphabets were effective in
discriminating decoys generated by gapless threading.  Here we showed
that similar conclusion can be drawn for statistical potential using
alphabet sets derived from natural clustering of residues, in
discriminating natives against more stringent compact decoy
conformations generated by off-lattice model \cite{ParkLevitt96_JMB}.
Our conclusion is also consistent with a recent study where it is
shown that much of the information in pairwise contact potential is
related to just a few variables such as hydropathy, charge, disulfide
bonding, and residue burial \cite{Cline02_Proteins}.

Although the alphabets we used have different number of residues, they
are all developed with one aspect in common, {\it i.e.}, the contacts
are all derived from the dual simplicial complexes, which provides a
faithful representation of the geometry.  This suggests that as long
as the same space filling pattern is conserved, the specific residue
types are not critical in many cases.  It seems that packing geometry
plays a very important role, but the specific residue types are often
replaceable.  This observation is consistent with experimental results
where it is well known that proteins are robust against many
mutations.

\paragraph{Edge Simplices and Tetrahedron Simplices.}
Pairwise alpha contact potential only considers the edge simplices or
1-simplices in the dual simplicial complex.  There have been several
studies of statistical potential based on 3-simplices or tetrahedra
\cite{Singh96_JCB,Zheng97,Carter01_JMB}.  In the work of Tropsha {\it
et al}, 3-simplices are obtained from unweighted Delaunay
triangulation \cite{Singh96_JCB,Zheng97}.  In these studies, all
residues are treated as balls of equal size located at C$_\alpha$ or
C$_\beta$ positions, and a cut-off distance is used to remove
tetrahedra that are considered to be too large.  Our approach is
different.  Our model is instead based on the weighted dual simplicial
complex, a different simplicial complex formed by a subset of the
simplices from the weighted Delaunay triangulation of all atoms in the
molecule.  The dual simplicial complex or the alpha shape allows
modeling at atomic level. Therefore, in our approach contacts can be
defined by the full side chain and main chain atoms.  Additionally,
atoms are assigned with appropriate non-uniform van der Waals radii
\cite{Tsai99_JMB}.  Finally, because the dual simplicial complex
reflects the precise contact geometry, we avoid the use of heuristic
cut-off thresholds necessary to eliminate a subset of the simplices
from the Delaunay triangulation. Our contacts represents accurately
geometry of the structure.  We discussed earlier the differences
between the alpha contact and contact by distance cut-off between
geometric centers of side chains.  It is conceivable that similar
difference will result between alpha contact and the approach
described in \cite{Singh96_JCB,Zheng97}.

\paragraph{Summary.}
In this work, we introduced a novel representation of protein
structures using edge simplices of the alpha shape, or the dual
simplicial complex of the protein structure.  By describing pairwise
contact interactions with simplicial edges, we developed alpha contact
potential based on the statistics of edge simplices. We also developed
a bootstrap model which provides confidence interval estimations,
including those of long range interactions.  We found that alpha
contact potential performs well in decoy structure discrimination.  By
comparing with alternative contact potential, we conclude that
geometric representation of contact interaction is important, but the
specific residue types are often interchangeable.

\section{Acknowledgment}
We thank Dr.\ Hui Lu for very helpful discussion. This work is
supported by funding from National Science Foundation CAREER
DBI0133856, DBI0078270 and MCB998008 and American Chemical
Society/Petroleum Research Fund 35616-G7.

\bibliography{nigms,svm,array,prop,pack,potential,prf,lattice,bioshape,liang,pair,career,design,pd02,cluster}
\bibliographystyle{unsrt}
\end{document}